\documentclass[10pt,journal,compsoc]{IEEEtran}

\ifCLASSOPTIONcompsoc
  \usepackage[nocompress]{cite}
\else
  \usepackage{cite}
\fi

\usepackage{authblk}

\ifCLASSINFOpdf
\else
 
\fi
\usepackage{amsmath}
\usepackage[normalem]{ulem}
\usepackage{subfig}
\usepackage{graphicx}
\usepackage[linesnumbered,ruled]{algorithm2e}
\usepackage{flushend}

\begin{document}
%

\title{Thread Progress Equalization: Dynamically Adaptive Power and Performance Optimization of Multi-threaded Applications}

\author[1]{Yatish Turakhia}
\author[2]{Guangshuo Liu}
\author[3]{Siddharth Garg}
\author[2]{Diana Marculescu}
\affil[1]{Department of Electrical Engineering, Stanford University}
\affil[2]{Department of Electrical and Computer Engineering, Carnegie Mellon University}
\affil[3] {Department of Electrical and Computer Engineering, New York
University }



\IEEEtitleabstractindextext{%
\begin{abstract}
Dynamically adaptive 
multi-core architectures have been proposed 
as an effective solution to optimize performance for peak power constrained 
processors.
In processors, the micro-architectural parameters or voltage/frequency 
of each core to be changed at run-time, thus providing a range of power/performance operating points for each core. 
In this paper, we propose Thread Progress Equalization (TPEq), 
a run-time mechanism for power constrained performance maximization of multithreaded applications running on dynamically adaptive multicore processors. Compared to existing approaches, TPEq (i) identifies and addresses two primary sources of inter-thread heterogeneity in multithreaded applications, (ii) determines the optimal core configurations in polynomial time with respect to the number of cores and configurations, and (iii) requires no modifications in the user-level source code. 
Our experimental evaluations demonstrate that TPEq outperforms state-of-the-art run-time power/performance optimization techniques proposed in literature for dynamically adaptive multicores by up to $23\%$.
\end{abstract}

}

\maketitle

\IEEEdisplaynontitleabstractindextext

\IEEEpeerreviewmaketitle

\ifCLASSOPTIONcompsoc
\IEEEraisesectionheading{\section{Introduction}\label{sec:introduction}}
\else
\section{Introduction}
\label{sec:introduction}
\fi

\IEEEPARstart{T}{echnology} scaling has enabled greater integration because of reduced transistor dimensions. 
Microprocessor designers have exploited the greater 
transistor budget to provision an increasing number of 
processing cores on the chip, effectively using thread-level 
parallelism to increase performance.
However, the 
power consumption per transistor has not been scaling commensurately with transistor dimensions~\cite{Esmaeilzadeh:2011:DSE:2000064.2000108}. 
This problem is compounded by the so-called ``power wall," a hard limit on the maximum power that a chip can draw. 
A critical challenge, in this context, is to devise techniques that maximize performance within a power budget. One solution to this problem is 
fine-grained, \textit{dynamic adaptation} at run-time. Broadly speaking, dynamic adaptation refers to the ability to dynamically distribute the available power budget amongst the cores on a multicore processor.


The traditional approach for fine-grained
dynamic adaptation is based on 
dynamic voltage and frequency scaling (DVFS), in which 
the voltage and frequency of each core (or group thereof) can be adjusted dynamically, providing a range of power/performance
operating points.
Moreover, recent 
work has advocated the use of micro-architectural
adaptation, in which the micro-architectural configuration
of each core to be adjusted dynamically (issue width, re-order buffer
size and cache capacity etc.), which is particularly effective in the 
context of ``dark silicon" era~\cite{Esmaeilzadeh:2011:DSE:2000064.2000108} where the power budget is 
constrained but transistors are abundant. 
Although the techniques proposed in this paper are described in the context of micro-architectural 
adaptation, 
they are equally applicable for DVFS  
and we 
provide experimental results for both dynamic adaptation techniques. 

To perform fine-grained dynamic 
adaptation, the operating system has to solve a challenging global optimization problem, i.e., how 
to determine the configuration of each core so as to maximize performance within the power budget. 
The problem is challenging for three reasons: (i) the solution must scale efficiently to multicore systems that have tens or even hundreds of cores; (ii) there is a
complex relationship between core configurations and the corresponding power/performance of the thread running on the core (this is particularly true for micro-architectural adaptation); 
and, (iii) for multithreaded applications, there is no direct performance metric to maximize, i.e., it is unclear how speeding up a single thread will affect the 
performance of the application as a whole.
These are the challenges that we address in this paper.

For sequential (i.e., single-threaded) applications, instructions per second (IPS) is a clear, measurable indicator of performance.
Moreover, for multiprogrammed workloads,
the IPS summed over all threads indicates net throughput, and is a commonly used performance metric~\cite{maxbips,herbert2007,liu2013}. 
However, for multithreaded applications, the sum of IPS metric can be a poor indicator of performance. For example, a thread that is spinning on a lock or waiting at a barrier might execute user-mode synchronization instructions, but these do \textit{not} correspond to useful work. 
The problem is heightened by the fact that programmers exploit parallelism in different ways --- for example, using data-level parallelism with barrier synchronization, or task-level parallelism with local/global task queues and static/dynamic load balancing. 

\begin{figure*}
        \centering
        \subfloat[][IPC heterogeneity only.]{
                \includegraphics[width=0.4\textwidth]{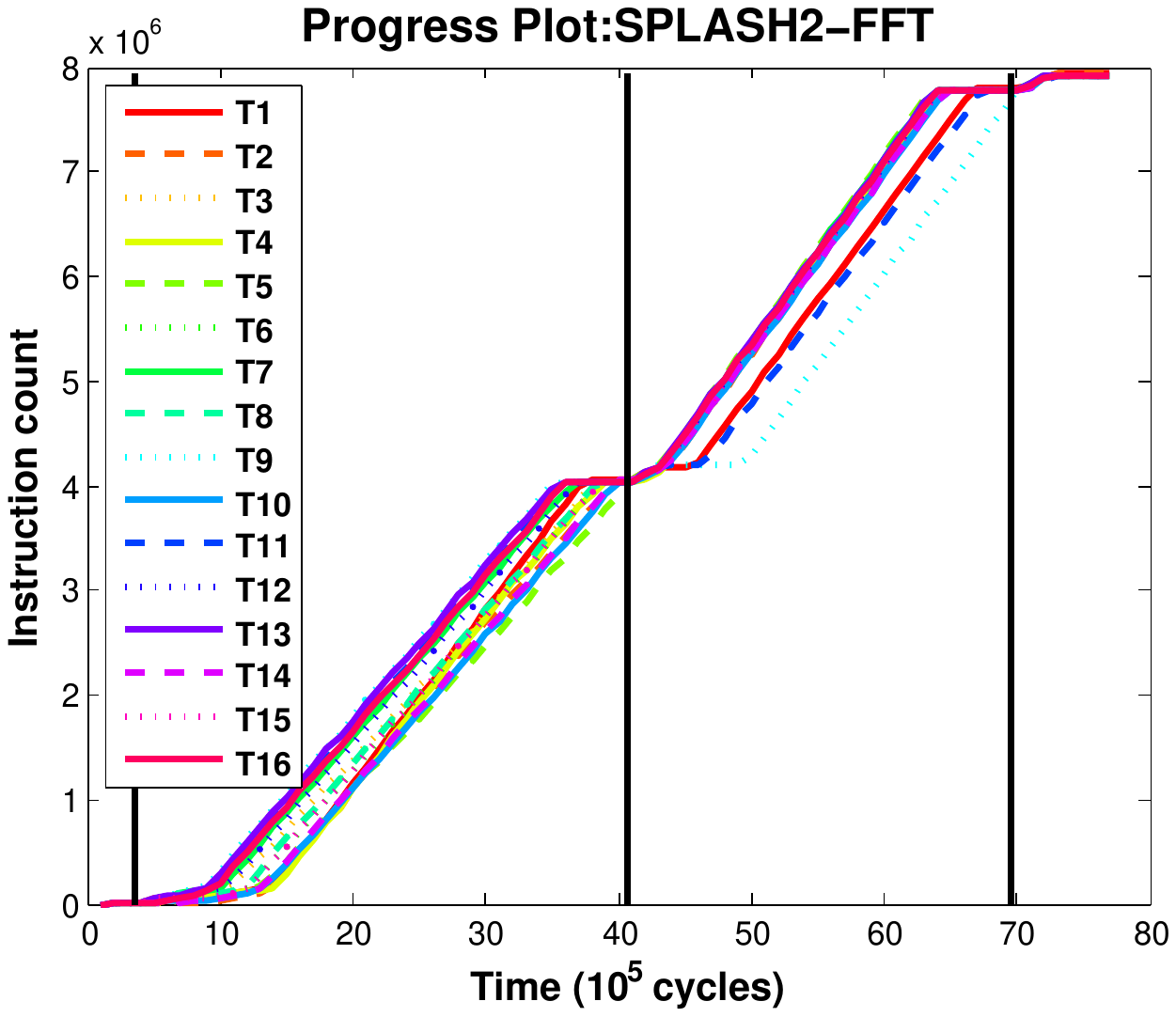}
                \label{fi:ipc}
        }
        \subfloat[][IPC and instruction count heterogeneity.]{
                \includegraphics[width=0.415\textwidth]{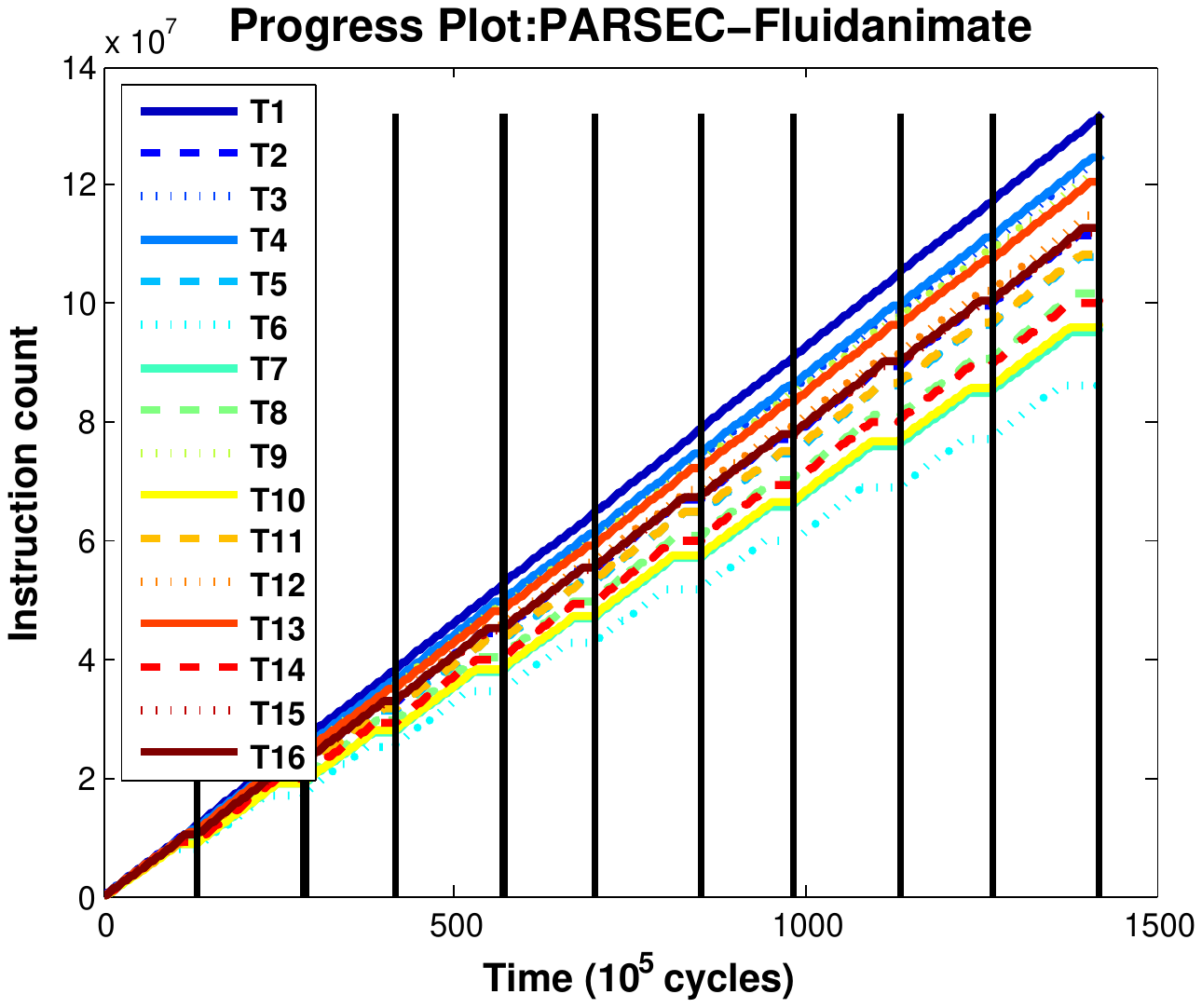}
                \label{fi:instr}
        }
        \caption{Progress plots for the FFT and Water.Nsquared (SPLASH-2) benchmarks with 16 threads on a 16-core architecture. The solid vertical lines indicate barriers. Slope corresponds to IPS of that thread and hence the flat regions indicate time periods when the thread is stalled waiting for lagging threads to arrive.}\label{fi:het-example}
\end{figure*}

\noindent \textit{Key Contributions} In this paper, we propose Thread Progress Equalization (TPEq), a run-time mechanism to maximize performance within a power budget for multithreaded applications running on multicore processors with per-core dynamic adaptation. The design of TPEq is motivated by multithreaded applications that make frequent use of barrier synchronization, but also generalizes, as we later discuss, to other models of parallelism.

We start with the observation that, to best utilize the available power budget, all threads that are expected
to synchronize on a barrier should arrive at the barrier at the same time. If this is not the case, 
\textit{early} threads (threads that arrive at a barrier earlier than others) can be slowed down and the power saved by doing so can be allocated to speed-up \textit{lagging} threads (threads that arrive at a barrier later than others). In this context, a natural question is why threads arrive at barriers at different times.

Empirically, we have observed two fundamental reasons for differences in the times at which threads arrive at barriers. First, even if each thread executes exactly the same sequence of instructions, threads can have different instructions per cycle (IPC) counts. For example, the sequence of data accesses that one thread makes can have less spatial locality than another thread's accesses, resulting in more cache misses and lower IPC for the first thread. We refer to this as \textit{IPC heterogeneity}. Second, each thread might execute a different number of instructions until it reaches a barrier. This is because the threads need not be inherently load balanced and depending on the input data, each thread can follow a different control flow path until it arrives at the barrier. 
We refer to this as \textit{instruction count heterogeneity}. 

Figure \ref{fi:het-example} shows an example of two benchmark applications, FFT and Water.Nsq (SPLASH-2~\cite{splash}), executing on a homogeneous multicore processor. FFT exhibits IPC heterogeneity but no instruction count heterogeneity, i.e., each thread executes exactly the same number of instructions between barriers. Water.Nsq exhibits both IPC heterogeneity, evident from different slopes of threads in progress plot, \textit{and} instruction count heterogeneity. Note that over the entire length of the application, thread T16 executes more than $1.15\times$ the number of instructions compared to thread T7. 

The goal of TPEq is to dynamically optimize the configuration of each core/thread such that each thread reaches the barrier at the same time by simultaneously accounting for \textit{both} IPC and instruction count heterogeneity
The design of TPEq is based on two components that operate synergistically:
\begin{itemize}
\item \textbf{TPEq Optimizer:} Given an oracle that can predict (i) the IPC
and power consumption of each thread for every core configuration, and (ii) the total number of instructions the thread must execute until the next barrier, we propose
and an efficient \textit{polynomial-time} algorithm that \textit{optimally} determines the core configuration for each thread to maximize application performance 
under power constraints.
\item \textbf{TPEq Predictors:} As input to the TPEq 
Optimizer, we implement accurate run-time predictors for (a) IPC and power consumption of a thread for different core configurations, and (b) the number of instructions each thread executes between barriers.  
\end{itemize}

TPEq is evaluated in the context of the Flicker~\cite{flicker} architecture, a recently proposed multicore processor design that supports dynamic adaptation of the micro-architectural parameters of each core. We compare TPEq to a number of existing techniques for power/performance optimization of multithreaded applications. 
\\ \\
\noindent \textbf{Distinguishing Features of TPEq:}
Compared to existing state-of-the-art approaches, TPEq has the following distinguishing features: (i) TPEq holistically accounts for both IPC and instruction count heterogeneity, 
while a number of other approaches only address one or the other; (ii) TPEq enables fine-grained adaptation for multicore processors where each core has multiple configurations; (iii) the TPEq optimizer provides optimal solutions in polynomial time, as opposed to other fine-grained optimization techniques that solve NP-hard problems and cannot achieve optimal results in less than exponential time; (iv) TPEq requires no software annotations or programmer specified progress metrics; and (v) TPEq generalizes to multithreaded applications that exploit different models of parallelization, including barrier synchronization, pipeline parallel and thread pool models with dynamic load balancing.

\section{TPEq Design and Implementation}

\begin{figure}
\centering
\includegraphics[width=0.35\textwidth]{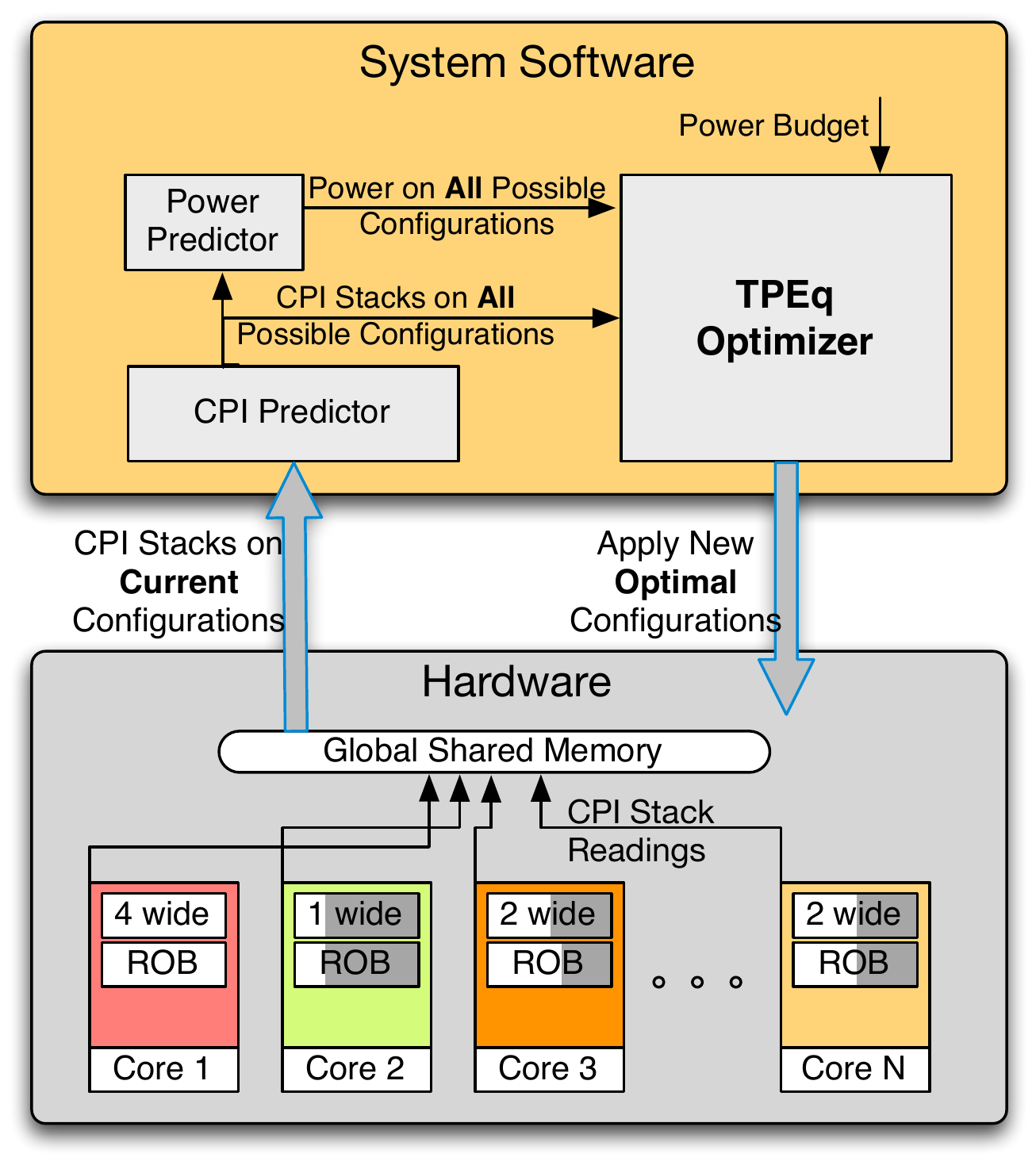}
\caption{Overview of TPEq approach on a dynamically adaptive multicore processor. 
}
\label{fi:overview}
\end{figure}

Figure~\ref{fi:overview} shows an overview of the design of TPEq. 
The hardware platform consists of a dynamically adaptive multi-core architecture 
where, for example, each core can have a 
variable ROB size and the fetch width.
In general, we will assume that each of the $N$ cores can be set in one of $M$ different \textit{configurations} as described in Table~\ref{tab:configuration}. 
In its current implementation, TPEq assumes that the number of threads equals the number of cores, and a static mapping of threads to cores~\cite{equal}. 
We believe TPEq can be extended to the case where there
are more threads than cores~\cite{unequal}, but leave that as a topic for future work.  

The TPEq run-time system consists of two components. The TPEq predictors monitor on-chip performance counters and predict the future application characteristics. The predictions are passed on to the TPEq optimizer, which determines the optimal configuration of each core so as
to maximize overall system performance within a power budget. We now describe the design and implementation of TPEq.

 
\subsection{TPEq Optimizer}

The TPEq optimizer is at the heart of TPEq approach. Although, in practice, the optimizer takes inputs from the TPEq predictor, we will discuss the optimizer
in the context of an oracle that provides the optimizer with perfect information and relax this assumption later.

To understand how the optimizer works, assume that we begin at the time instant when $N$ threads exit a barrier and start making progress towards the next barrier. The optimal configuration for each core/thread needs to be decided for the interval between these two successive barriers. Assume that an oracle provides access to the following information:

\begin{itemize}
\item The number of instructions each thread executes until it enters the next barrier is in ratio $w(1):w(2):\ldots:w(N)$. Note that $w(1), w(2), \ldots , w(N)$ can be absolute instruction counts, but we only require the number of instructions each thread executes relative to other threads.  
\item The CPI of thread $i$ ($1 \leq i \leq N$) when it executes on a core with configuration $j$ ($1 \leq j \leq M$) is $CPI(i,j)$, and the corresponding power dissipation is $P(i,j)$. We assume, for now, that for a given core configuration, the CPI and power dissipation of each thread do not change with time, at least until it reaches the next barrier. This assumption is relaxed later.
\end{itemize}

Under the assumptions above, TPEq tries to assign a configuration to each core/thread so as to stay within power budget $P_{budget}$, while minimizing the time taken by the most lagging thread 
to reach the next barrier. A key contribution of our work is an algorithm that \textit{optimally} solves this problem in $\mathcal{O}(MN \log N)$ time.

The algorithm works as follows: TPEq starts by setting all cores to the configuration that consumes the least power and determines the identity of the
\textit{most lagging} thread for this setting, i.e., the thread that would reach the barrier
last. For thread $i$, the number of clock cycles required to reach the barrier when executing on configuration $j$ would be
$w(i)CPI(i,j)$. We define the \textit{progress} of this thread as:
$$ progress(i) = \frac{1}{w(i)CPI(i,j)}$$ 
to capture the intuition that larger
values of ``progress" are better.

The configuration of the most lagging thread is then moved up to the next level\footnote{Without loss of generality, the configurations are, by convention, sorted in ascending order of power consumption. 
Also, we limit the search to \textit{Pareto optimal} configurations, by simply discarding ones where increasing power does not lead to increased performance.}, and the new most lagging thread is determined. The core configuration for this new most lagging thread is now moved up by one level, and so on.
This continues until there is no core whose configuration can be increased to the next level without violating the power budget. 
The resulting core configurations are optimal in terms of total execution time and are then updated in hardware.
Algorithm \ref{algo:TPEq} is a formal description of this optimization procedure.

\begin{algorithm}
\caption{ TPEq Optimization Procedure } 
\label{algo:TPEq} 
  $P_{tot} \gets 0$\;
  \tcp{Init. all threads to lowest core config.}
  \For{$i \in [1,N]$}{
    $c(i) \gets 1$\; 
    $P_{tot} \gets P_{tot} + P(i,c(i))$\;
    \label{line:progress metric}
    $progress(i) \gets \frac{1}{{w(i)}{CPI(i,c(i))}}$\;
  }
  \While{$P_{tot} \leq P_{budget}$}{
    \tcp{Determine lagging thread $l$}
    $l \gets arg\,min_{i \in [1,N], c(i) < M} \left\{ \frac{1}{w(i) CPI(i,c(i))} \right \}$\;
    \label{line:identify_lagging}
    \tcp{If no such thread exists}
    \If{$l = { \emptyset }$ }{break\;}
    \tcp{Increase core configuration of lagging thread}
    $c(l) \gets c(l)+1$\;
    \label{line:increase_configuration}
    \tcp{Update progress and power}
    $progress(l) \gets \frac{1}{w(l)CPI(l,c(l))} $\;
    $P_{tot} \gets P_{tot} - P(l,c(l)-1) + P(l,c(l))$\;
  }
  \tcp{Return optimal core configurations}
  \Return $c$\;
  \vspace{0.1in}
\end{algorithm}

We now provide a formal proof of optimality for this algorithm below. 

\noindent \textbf{Proof of optimality (by contradiction):} 
Let \emph{C} = $\textless c(1)\ c(2)\ ...\  c(N)\textgreater$ be the TPEq configuration vector of cores for $N$ threads, such that $c(1)$ corresponds to the core configuration of thread $1$, $c(2)$ corresponds to the core configuration of thread $2$, and so on.  Let $P_{tot}$ be the total power consumption with configuration vector \emph{C}, such that $P_{tot} \leq P_{budget}$. Let $progress(i, c(i))$ denote the progress of thread $i$ with core configuration $c(i)$ and $min\_progress(C)$ denote the progress of the most lagging thread with configuration vector $C$. Since only Pareto optimal configurations are considered, $progress(i, c(i)) > progress(i, c^*(i)) \implies P(i, c(i)) > P(i, c^*(i))$. Now assume a better configuration vector \emph{C*}=$\textless c^*(1)\ c^*(2)\ ...\ c^*(N)\textgreater$  with total power $P_{tot}^*$ within same power budget exists i.e $min\_progress(C^*) > min\_progress(C)$ and $P_{tot}^*\leq P_{budget}$. 

First, consider a case in which TPEq does not assign a configuration to any thread that is larger than the optimal configuration i.e $c(i) \leq c^*(i)\ \forall i \in [1,N]$. This also implies that $P_{tot} \leq P_{tot}^*$. Without loss of generality, assume that first $K$ threads have strictly larger core configurations in the optimal assignment i.e $c(i) < c^*(i)\ \forall i \in [1,K]$ and the remaining threads have same configurations as TPEq $c(i) = c^*(i)\ \forall i \in [K+1,N]$. If the most lagging thread $l$ for configuration $C$ was in the first $K$ threads, the algorithm~\ref{algo:TPEq} would not terminate as it is possible to increase core configuration of thread $l$ to $c^*(l)$ and remain within total power $P^*_{tot}$ (and therefore, $P_{budget}$). If $l \in [K+1,N]$, $c(l)=c^*(l)$ and therefore, $min\_progress(C) \geq min\_progress(C^*)$. This is a contradiction. 

Next, consider a case in which TPEq assigns a configuration to a thread $j$ that is larger than the optimal configuration i.e. $c(j) > c^*(j)$. This implies $progress(j, c(j))>progress(j,c^*(j))$. But since TPEq only accelerates the most lagging thread in each iteration and since TPEq assigned thread $j$ to $c(j)$, which is larger than $c^*(j)$, $min\_progress(C) \geq progress(j,c(j))$. This implies $min\_progress(C) \geq progress(j,c^*(j)) \geq min\_progress(C^*)$. Again, a contradiction. Therefore, TPEq configuration is optimal.


\noindent \textbf{Min-heap Based Implementation:} A naive implementation of the 
TPEq optimization algorithm (Algorithm~\ref{algo:TPEq}) 
would use a linear array to store the 
progress metric of each thread 
(line~\ref{line:increase_configuration}), 
resulting in a $\mathcal{O}(MN^{2})$ time complexity. 
However, note that 
in each iteration of Algorithm~\ref{algo:TPEq}, we only update the progress 
of the currently slowest thread, i.e., least progress. 
Based on this observation, 
we propose an improved implementation of Algorithm~\ref{algo:TPEq} 
with $\mathcal{O}(MN \log N)$ 
that stores
the progress metric of each thread in a \textit{min-heap} data-structure.
A min-heap is a binary tree where the data in each node is less than 
or equal to its children.

In the proposed implementation, 
setting up the min-heap data structure takes $\mathcal{O}(\log N)$ time 
(line \ref{line:progress metric}), determining the currently most lagging 
thread takes $\mathcal{O}(1)$ time (line \ref{line:identify_lagging}), and 
updating the progress metric of the lagging thread and reinserting it back 
into the heap takes $\mathcal{O}(\log N)$ 
time (line \ref{line:increase_configuration}).
Finally, the outermost \textit{while} loop iterates at most 
$MN$, resulting in a time complexity of $\mathcal{O}(MN \log N)$. 

\noindent \textbf{Epoch Length}
In practice, the TPEq optimization routine is called once every \textit{epoch} in order to address fast variations in thread characteristics. The epoch length ($\mathcal{E}$, measured in number of clock cycles) is configurable. The epoch length should be short enough to quickly adapt to CPI and power variations, but is practically limited by the computational overhead of the optimization procedure. In this context, the polynomial time complexity of the TPEq optimizer, which counts for less than $1\%$ run-time overhead for a $1$ $ms$ epoch, enables the use of relatively fine-grained temporal adaptation that would be otherwise impractical.

Note that since epochs are not necessarily synchronized with barriers, in practice we need a slightly updated progress metric from the one used in Algorithm~\ref{algo:TPEq}. Therefore, the progress of a thread is measured in terms of its \textit{predicted} progress by the end of the current epoch:
\begin{equation} \label{eq:progress}
progress(i) = \frac{instrCount(i)}{w(i)} + \frac{\mathcal{E}}{w(i) CPI(i,c(i))},
\end{equation}
where the first time represents progress made so far and the second term represents predicted progress in the next epoch.

\subsection{TPEq Predictors}\label{TPEq_pred}

In the previous sub-section we assumed that the TPEq optimizer 
has oracular knowledge of the relative instruction counts of the threads. 
In practice, the TPEq predictors determine these 
values at run-time for each thread immediately 
after a synchronization related stall. 
TPEq also requires predictions for CPI and power consumption of each thread for every core configuration once every \textit{epoch}, i.e., in synchrony with the TPEq optimization procedure. 

\subsubsection{Relative Instruction Count Prediction} 
We start by describing the relative instruction count predictor. The instruction count predictor predicts the number of instructions each thread executes relative to other threads. Our predictor
is based on the observation that \textit{the number of instructions each thread executes between barriers, relative to other threads, remains the same}. This motivates the use of a history based predictor to predict
relative instruction counts.

Intuitively, we note that the difference in relative instruction counts of several multithreaded workloads arise as a result of imbalance in the amount of computing for a thread, which persists across several barriers. 
Singh et al.~\cite{Singh1995118} were perhaps the first to 
qualitatively 
observe the locality in the data distribution in threads 
across successive barriers in many of our benchmark algorithms and 
provide insights into this characteristic. They noted that 
successive barriers correspond to very small ``time-steps" 
in the physical world, and that the characteristics of the
physical world change slowly with time. Hence, the amount 
of work to be performed by a thread in one time-step, is a good predictor
for the amount of work in the next time-step. 
In the progress plot for Water.Nsq (see Figure~\ref{fi:instr}), for instance,
the number of water molecules per thread remain nearly constant across barriers. 
Consequently, thread T16 (T7), with most (least) number of water molecules, always executes the most (least) instructions in any inter-barrier interval.

\begin{figure}
\centering
\includegraphics[width=0.4\textwidth]{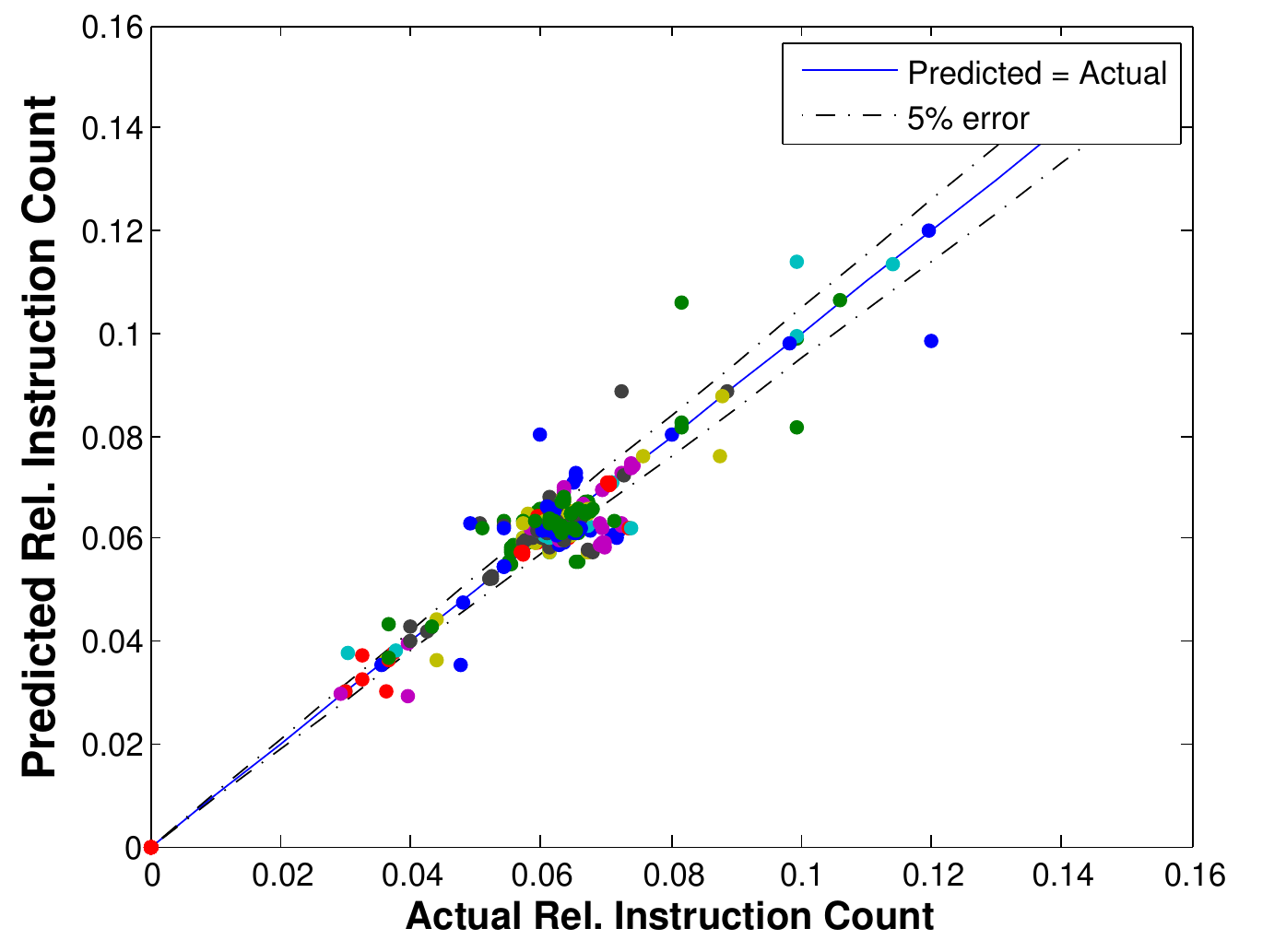}
\caption{Scatter plot of predicted and actual relative instruction counts between successive barriers for different benchmarks.
}
\label{fi:scatter_pred}
\end{figure}

Quantitatively, we have verified this trend 
over all barrier synchronization based benchmarks in the SPLASH-2, PARSEC and 
Phoenix benchmarks suites (see Table~\ref{tab:workloads} for more details) that we experimented with.
In particular, 
Figure~\ref{fi:scatter_pred} shows 
the a scatter plot of relative instruction counts in barrier phase $t+1$ 
versus the relative instruction counts of threads in barrier phase $t$
across all benchmarks with instruction count heterogeneity (coded in different colors).
The mean absolute relative error using a last-value-predictor
for relative instruction counts was found to be only $4.2\%$. 
Liu et al.~\cite{liu2005exploiting} have observed
similar locality behaviour across the outermost loops of the SpecOMP parallel applications, and  
use last-value prediction to perform voltage/frequency scaling for each thread. However, there 
are significant differences between their work and ours and these are discussed in Section~\ref{related}.



\noindent \textbf{Implementation Details} 
The TPEq relative instruction count predictor keeps a running count of the number of user mode instructions executed by a thread. The relative instruction count, $w(i)$, of each thread is updated with its running count at the end of 
\textit{any} synchronization related stall. This technique
is simple, requires no synchronization between threads to detect barriers
and avoids the need
for the user to indicate when barriers occur. 

In our scheme, if an application has only barrier related synchronization, 
all threads will update as they exit the barriers. 
In addition, the weight in any inter-barrier interval will 
automatically correspond to the average number of instructions per barrier executed by the thread so far. 
In fact, TPEq does not distinguish barrier related stalls from other synchronization stalls 
such as ones due to critical sections. 
For
one, barriers can be implemented
using other synchronization primitives as well, 
including locks~\cite{mellor1991algorithms}, which we would like to
capture. As well, taking into account \textit{all} synchronization related stalls introduces certain advantages that will be discussed in later sections.

Any synchronization related stall detection mechanism can be used
to determine when these stalls occur. Hardware based thread progress
metrics have been proposed~\cite{li2006spin} to detect threads that are spinning on locks or waiting at barriers. 
These solutions are the most general but suffer from false positives and true negatives, resulting in incorrect optimization decisions. 
Alternatively, software based solutions
can be implemented, either using programmer or compiler inserted annotations, 
or by modifying the threading library and OS synchronization primitives~\cite{cs}. 
We adopt the latter approach. As in ~\cite{cs}, 
we detect scenarios in which the threads are stalled due to synchronization 
and update our predictors with current relative instruction counts while exiting the stall. 


\vspace{0.05in}



\label{sec:prediction}
\subsubsection{CPI and Power Prediction} 
\begin{figure}
\centering
\includegraphics[width=0.35\textwidth]{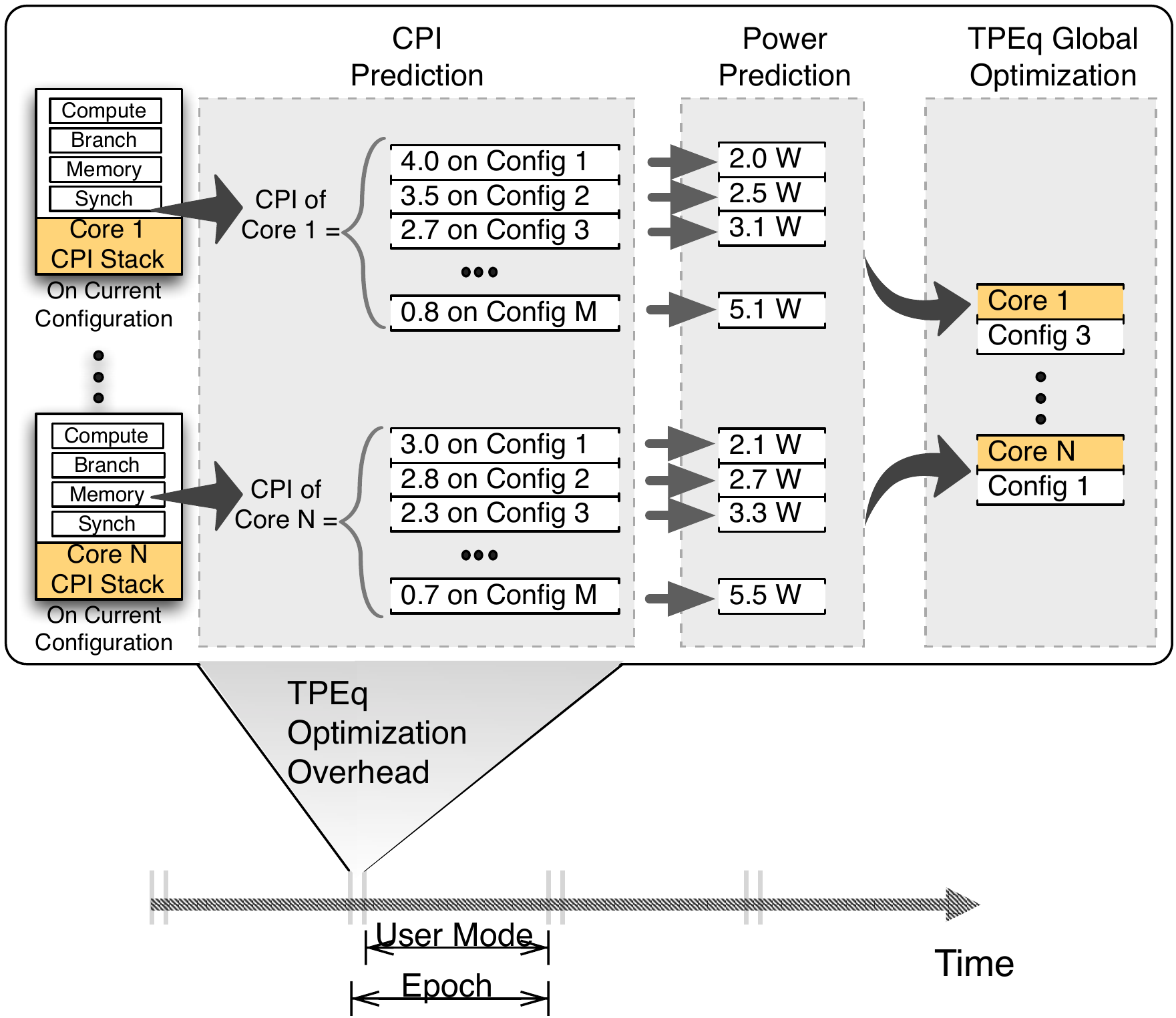}
\caption{CPI and power prediction overview. A detailed discussion of the TPEq predictors is in Section~\ref{sec:prediction}.
}
\label{fi:prediction}
\end{figure}

We now describe the CPI and power predictors that we use in TPEq which, 
as shown in Figure~\ref{fi:prediction}, 
are called once every epoch.

Let $CPI_{t}(i,j)$ be the CPI of thread $i$ on core configuration $j$ in epoch $t$. The goal of the CPI predictor is to determine $CPI_{t+1}(i,j)$ for all $j \in [1,M]$. 
Duesterwald {\em et al.}~\cite{dwarka} have shown that for predicting the CPI in the next epoch assuming the same core configuration, i.e., characterizing temporal variability in CPI, last-value predictors perform on a par with exponentially-weighted mean, table-based  and cross-metric predictors. 
The accuracy of last-value predictors improves for shorter prediction epochs. We choose to use a last-value predictor in TPEq because of its simplicity, and because we are able to afford relatively short epoch lengths. The last-value predictor simply implements:
$$CPI_{t+1}(i,j) = CPI_{t}(i,j).$$

To predict $CPI_{t+1}(i,k)$ for all $k \neq j$ given $CPI_{t+1}(i,j)$ , we need an approach that predicts the performance on one core type given performance on another core type. 
%
%
For this, TPEq uses CPI stack information measured using hardware counters broken down into four components: compute CPI (base CPI in the absence of miss events), memory CPI (cycles lost due to misses in the memory hierarchy), branch CPI (cycles lost due to branch misprediction) and synchronization CPI (cycles lost due to stalls on synchronization instructions).

With these measurements on configuration $j$, we predict the CPI on configuration $k$ using a linear predictor as follows: 
\begin{eqnarray}
CPI_t(i,k) = \alpha^{comp}_{jk}CPI^{comp}_{t}(i,j) + \alpha^{mem}_{jk}CPI^{mem}_{t}(i,j) \nonumber \\
+ \alpha^{branch}_{jk}CPI^{branch}_{t}(i,j)   + \alpha^{synch}_{jk}CPI^{synch}_{t}(i,j). \nonumber
\end{eqnarray}
The pairwise $\alpha^{*}_{jk}$ parameters, one for every pair of core configurations, are learned offline using training data obtained from a set of representative benchmarks and stored for online use. Note that the learned parameters are not benchmark specific and depend only on the core configurations.

A similar linear predictor that utilizes CPI components was proposed by Lukefahr et al.~\cite{composite}, although only for big-little core configurations. 
Another CPI predictor is PIE~\cite{pie}, which makes use of information collected using hardware counters including the total CPI, CPI of memory instructions, misses per instruction (MPI), and data dependencies between instructions. However, 
PIE has been proposed for CPI 
prediction between small in-order and large out-of-order cores, while TPEq also requires 
predictions between different out-of-order core configurations and also faces the challenge of predicting over future epoch. 
Furthermore, since training the TPEq predictor is automated and data-driven, 
it is easy to deploy for a large number of core configurations.  


We note that existing processors such as the Intel Pentium 4~\cite{intelp4} and the IBM POWER5~\cite{ibmp5} have built-in hardware support for performance counters that measure CPI components. In addition, Eyerman et al.~\cite{eyerman2006} have proposed a performance counter architecture that further improves upon the accuracy of these commercial implementations with similar hardware complexity. Their approach provides very accurate estimates of the CPI stack components with only $2\%$ average absolute error.

The TPEq power predictor uses the predicted CPI values for each core 
configuration (as described above) to predict their power consumption.
This is based on previous work which indicates that
that CPI (or IPC) is highly correlated with 
power consumption~\cite{contreras2005power,bircher}; 
for instance, 
Bircher and John report on average only $3\%$ error when compared to 
measured CPU power~\cite{bircher}. 
Indeed, we empirically verified 
that incorporating more fine-grained data like the individual CPI stack 
components did not improve the accuracy of power prediction 
significantly. 
However, we did observe that moving from a simple linear 
predictor to a quadratic 
model did improve accuracy. 
Thus, the TPEq power predictor predicts the power consumption for different core types as follows:
$$P(i,j) = \beta_{0,j} + \frac{\beta_{1,j}} { CPI(i,j)} + \frac{\beta_{2,j}}{ {CPI(i,j)}^{2} }$$
where  $\beta_{0,j}$,  $\beta_{1,j}$, and  $\beta_{2,j}$ are fixed parameters that are learned for each core type offline and stored for online use.


\subsection{Implementation Details}\label{imp}

The TPEq optimization and prediction routines are implemented in software. The primary
hardware overhead that TPEq introduces is the hardware required to track the CPI stack components. 
As mentioned before, existing commercially available processors such as the Intel Pentium 4 and IBM POWER5
already have hardware performance counters to measure CPI stack components.

Based on the design proposed by Eyerman et al.~\cite{eyerman2006}, we estimate the hardware overhead for TPEq as follows: (i) one global 32-bit register and counter per CPI stack component (five registers/counters in all), (ii) a shared front-end miss table (sFMT) with as many rows as number of outstanding branches supported, an ROB ID and local branch misprediction penalty counter per-row, and a shared I-cache/TLB miss counter, (iii) a back-end miss counter for D-cache/D-TLB misses; and (iv) a long latency functional unit counter. The counters in (ii), (iii) and (iv) are all local counters and only need to count up to the maximum miss penalties for their respective events.

The TPEq prediction and optimization procedures are invoked by the OS in every epoch using an interrupt. The CPI stack values on each core are stored to shared memory, after which one core, designated as the leader, reads these values, performs CPI and power predictions and determines the optimal core configurations. All other cores are stalled in this period. Finally, the configuration of each core is updated based on the optimal configurations and control is passed back to user code. In the empirical results section, we quantify all the execution time overheads of the TPEq procedures.

\subsection{Comparative Analysis of TPEq}\label{comparitive}

To provide more insight into
our proposed approach, we compare TPEq qualitatively to three state-of-the-art approaches 
for maximizing the performance of multi-threaded applications.

%
%

\noindent \textbf{Criticality Stacks (CS):} Criticality Stacks~\cite{cs} is a recently proposed 
metric for thread criticality that measures the amount
of time in which a thread is active (not stalled due to synchronization)
in each epoch divided by the number of other threads active
in the same epoch. Intuitively, the most critical thread is one that
is active while all others are stalled. 



TPEq 
incorporates a notion of 
criticality similar to that of 
CS through the weights  $w(i)$. 
Threads that spend more (less) time stalled will 
have lower (higher) $w(i)$ values for TPEq, and 
similarly, lower (higher) criticality values in CS. 
The numbers will not be identical though, 
since the time spent in active state is weighted differently in the two approaches. 

Most importantly, CS is a 
coarse-grained optimization techniques, in that it only
accelerates the ``most-lagging" thread. 
In contrast, the TPEq optimizer performs 
fine-grained optimization based on the progress metric and weight of every thread, and is therefore able to best utilize the available power budget.

We note that CS is itself a generalization
of Age-based Scheduling~\cite{agets} (AGETS), in which the 
thread which has executed the least number of instructions relative to other threads is sped up on a faster core. Although we also 
implemented and experimented with AGETS~\cite{agets}, we 
found that CS outperformed AGETS across
all the benchmarks we studied, so we do not report any data for AGETS 
in this paper.



\noindent \textbf{MaxBIPS:} Maximizing sum-IPS~\cite{maxbips} is a commonly used (and intuitive) objective
for applications where the threads are independent --- multiprogrammed workloads, for example, 
or multithreaded benchmarks with dynamically load balanced task
queues and task stealing~\cite{herbert2007}. 
Like TPEq (and unlike CS), MaxBIPS can be used for fine-grained optimization of core configurations.
However, the primary 
problem with MaxBIPS in the context of 
multi-threaded benchmarks is that it has no notion of thread synchronization and does not 
take thread criticality into account.

\noindent \textbf{Bottleneck Identification and Scheduling:} Bottleneck Identification and Scheduling~\cite{bis} (BIS) annotates critical sections in the code and uses these annotations to determine and accelerate bottlenecks, i.e., performance critical threads, at run-time. As opposed to the previously discussed techniques, BIS does require access to source code, and, at least for a set of benchmarks evaluated, CS performs at least as well as BIS~\cite{cs}.

Nonetheless, we believe techniques such as BIS, and its updated version UBA~\cite{uba}, are orthogonal to and can be used in conjunction with TPEq. For example, threads that have BIS-based criticality greater than a threshold can be assigned to the highest core configuration, while the remaining $N-1$ cores configurations can be optimized using TPEq. We leave this as future work.


\begin{table}
\resizebox{0.49\textwidth}{!}
{
    \centering
    \begin{tabular}{|r||l|l|l|}
    \hline
    Configuration & Dispatch width & ROB size & Integer ALUs\\
    \hline
    1 & 1 & 16 & 1\\
    2 & 2 & 32 & 3\\
    3 & 2 & 64 & 3\\
    4 & 4 & 64 & 6\\
    5 & 4 & 128 & 6 \\
    \hline
    \multicolumn{4}{|l|}{Number of cores: 16, Number of threads: 16 (1 thread/core)}\\
    \multicolumn{4}{|l|}{Frequency: 3.5 GHz, Voltage: 1.00 V, 22nm Technology Node}\\
    \multicolumn{4}{|l|}{L1-I cache: 128 KB, write-back, 4-way, 4-cycle}\\
    \multicolumn{4}{|l|}{L1-D cache: 128 KB, write-back, 8-way, 4-cycle}\\
    \multicolumn{4}{|l|}{L2 cache: private 256 KB, write-back, 8-way, 8-cycle}\\
    \multicolumn{4}{|l|}{L3 cache: 8 MB shared/4 cores, write-back, 16-way, 30-cycle}\\
    \multicolumn{4}{|l|}{Cache coherence: directory-based MSI protocol}\\
    \multicolumn{4}{|l|}{Floating point units: 2, Complex ALUs: 1 }\\
    \hline
    \end{tabular}}
    \caption{Microarchitectural adaptation configurations}
    \label{tab:configuration}
\end{table}

%
\begin{table}
\resizebox{0.49\textwidth}{!}
{
    \centering
    \begin{tabular}{|r|l|l|l|l|l|}
    \hline
 & Conf. 1 & Conf. 2 & Conf. 3 & Conf. 4 & Conf. 5\\
    \hline
   IPC & 0.65 & 1.06 & 1.13 & 1.27 & 1.36 \\
    Power (W) & 3.93 & 5.43 &  5.56 & 6.51 & 6.69\\
    \hline
    \end{tabular}}
    \caption{Maximum IPC and power observed for different configurations using Swaptions.}
    \label{tab:IPC_power_range}
\end{table}


\section{Experimental Setup}

Our empirical evaluation of TPEq is based on the Sniper~\cite{sniper} multicore simulator for x86 processors.
We augment Sniper with our TPEq code and our patch that 
enables dynamic adaption of hardware parameters, 
including front-end pipeline 
width and reorder buffer (ROB) size, and scripts for the other state-of-the art techniques we compare against.
For power estimation, we use McPAT~\cite{mcpat}, which is seamlessly integrated with Sniper.

We model a processor with 
16 cores and an 80W power budget. 
The relevant core/uncore micro-architectural 
parameters are
shown in Table~\ref{tab:configuration}. 
Each core can pick from one of five different configurations 
which are also listed in Table~\ref{tab:configuration}.
We note that in our experiments, the issue queue 
and load-store queue are scaled automatically with ROB 
size, since in 
Sniper all 
three are governed by a single parameter 
``\textit{window size}"). 
Table~\ref{tab:IPC_power_range} shows the maximum observed IPC and power values over all epochs for different static core configurations using Swaptions benchmark.
Finally, in all experiments, the epoch length is set to $1$ $ms$ (3.5 million clock cycles at the baseline clock frequency of 3.5 GHz). 


\begin{table}
    \centering
    \begin{tabular}{ l | l | l }
    Category & Workload & Benchmark Suite\\
    \hline\hline
    Homogeneous (HO)& Blackscholes & PARSEC \\
    & Canneal & PARSEC\\
    & FFT & SPLASH-2\\
    & Ocean.cont & SPLASH-2\\
    & Radix & SPLASH-2\\
    & Streamcluster & PARSEC\\
    & Swaptions & PARSEC\\
    \hline
    Heterogeneous (HT)& Barnes & SPLASH-2 \\
    & Fluidanimate & PARSEC\\
    & LU.cont & SPLASH-2\\
    & LU.ncont & SPLASH-2\\
    & Water.nsq & SPLASH-2\\
    & Water.sp & SPLASH-2\\
    & Bodytrack & PARSEC \\
    & Kmeans & Phoenix\\

%
    \end{tabular}
    \caption{Barrier synchronization based benchmarks classified as either
    homogeneous or heterogeneous.}
    \label{tab:workloads}
\end{table}

\begin{table}
    \centering
    \begin{tabular}{ l | l | l }
    Category & Workload & Benchmark Suite\\
    \hline\hline
    Thread pool (TP)
    & Cholesky & SPLASH-2 \\
    & Radiosity & PARSEC\\
    \hline
    Pipeline Parallel (PP)& Dedup & PARSEC \\
    & Ferret & PARSEC\\
    \hline
    MapReduce (MR)& Histogram & Phoenix \\
    & Linear regression & Phoenix\\
    & Matrix multiply & Phoenix\\
    & String match & Phoenix\\
    & Word count & Phoenix\\
%
    \end{tabular}
    \caption{Benchmarks using alternative approaches to parallelization.}
    \label{tab:workloads-other}
\end{table}

\begin{figure*}
        \centering
        \subfloat[][TPEq compared to CS and MaxBIPS]{
                \includegraphics[width=0.49\textwidth]{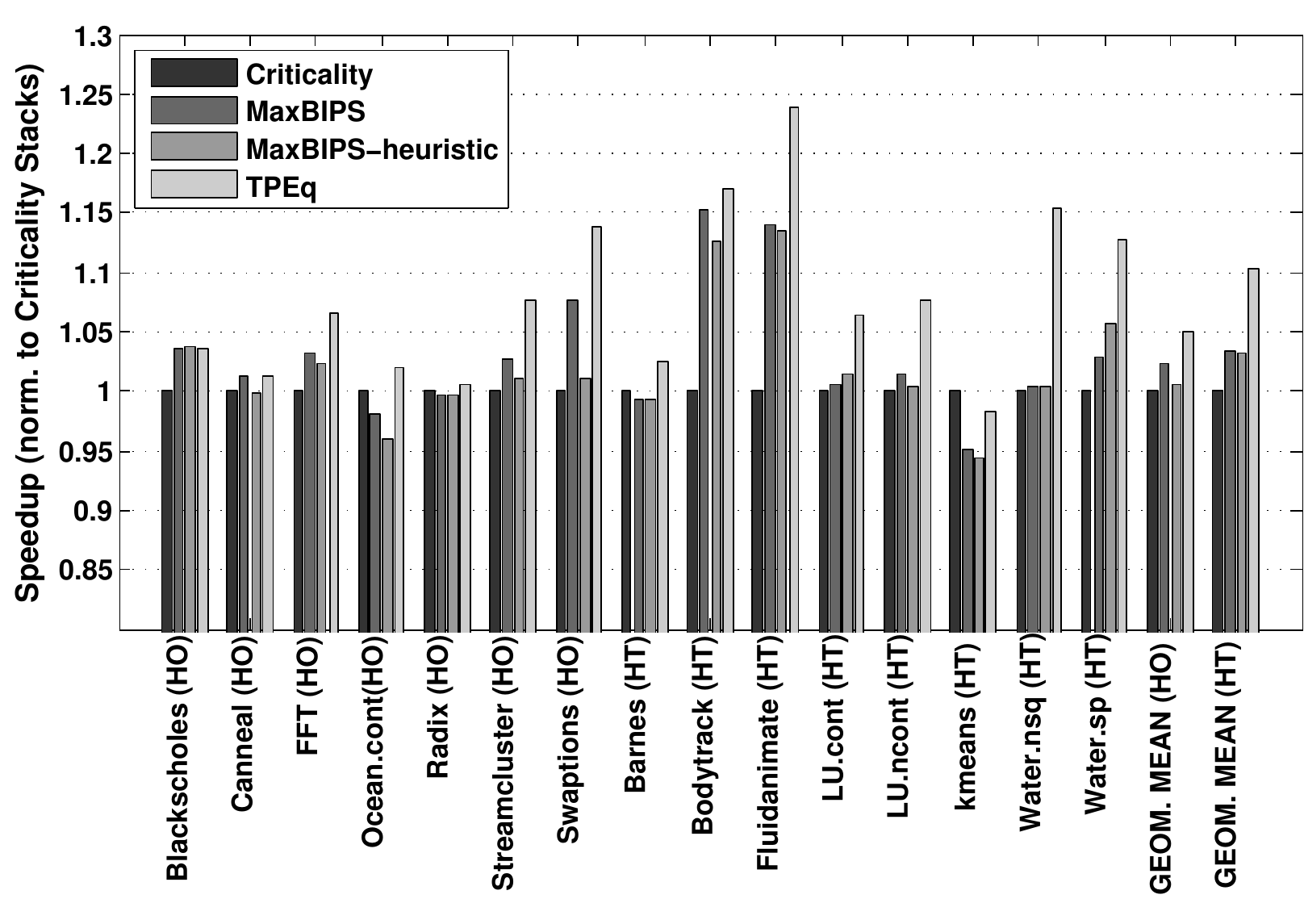}
                \label{fig:tpeq_compare_hom_het_a}
        }
        \subfloat[][Benefits of relative instruction count prediction (weights).]{
                \includegraphics[width=0.49\textwidth]{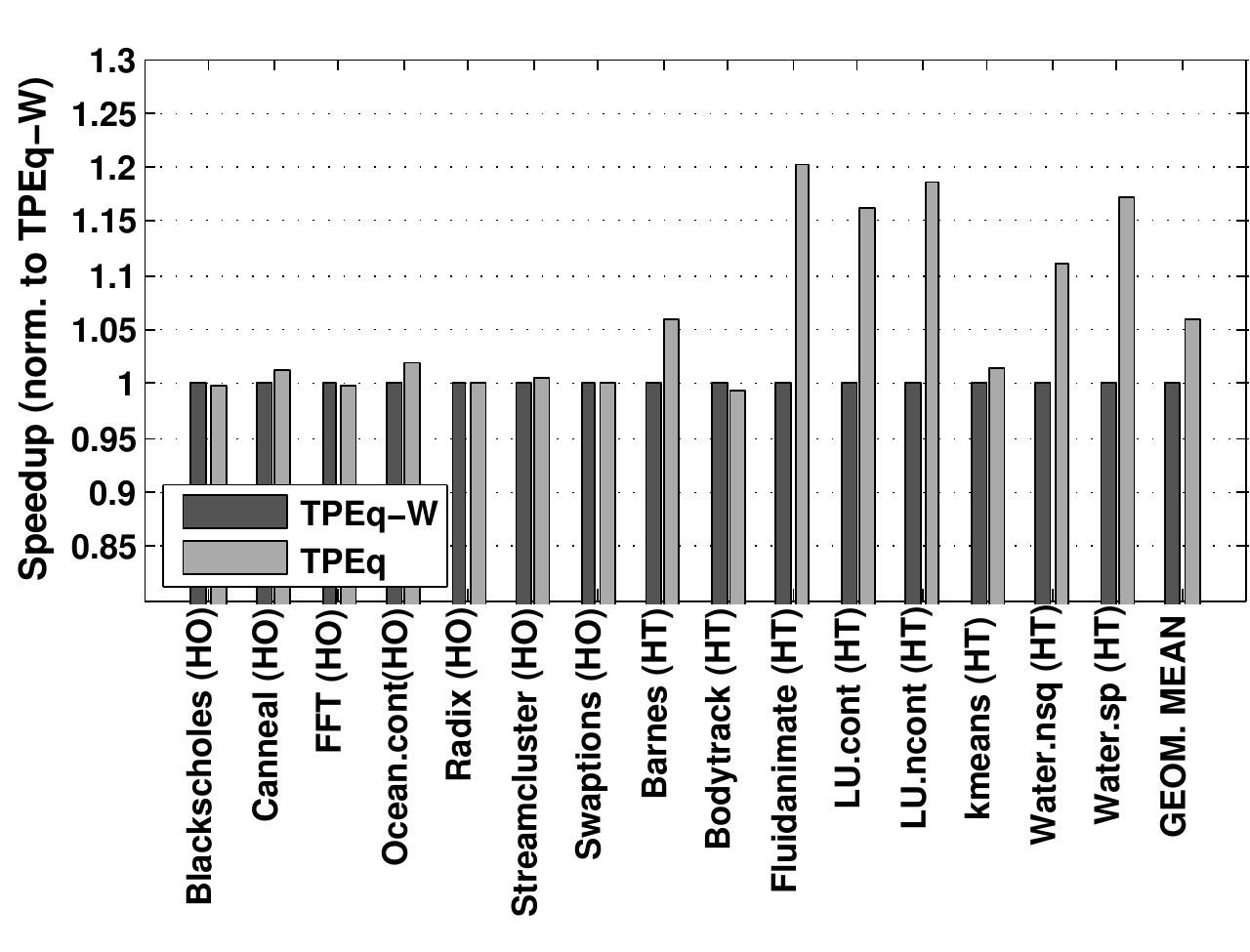}
                \label{fig:tpeq_compare_hom_het_b}
        }
        \caption{(a) Speed-up of TPEq and MaxBIPS using the execution time of CS as baseline. Also shown is MAXBIPS-heuristic. (b) Speed-up of TPEq with respect to TPEq without relative instruction count prediction (TPEq-W), i.e., where all the weights are set to one.}\label{fig:tpeq_compare_hom_het}
\end{figure*}

The workloads used in our experiments are multithreaded 
applications from the PARSEC~\cite{parsec}, SPLASH-2~\cite{splash} 
and Phoenix~\cite{ranger2007evaluating} benchmark suites. 
We have included 18 out of 22 benchmarks in SPLASH-2 and PARSEC 
combined, excluding only the ones which we had compilation or 
run-time issues with Sniper. 

Table~\ref{tab:workloads} shows the subset of the benchmarks that extensively
make use of barrier synchronization for parallelization. These are the benchmarks for which
we expect TPEq to perform the best, since it is designed keep barrier synchronization based
parallelism in mind. These benchmarks are further classified as: (i) homogeneous benchmarks: threads execute the same number of instructions between barriers; (ii) heterogeneous: threads execute the same number of instructions between barriers relative to each other.

Table~\ref{tab:workloads-other} show the remaining benchmarks that use other
types of parallelism. We classify 
these as follows: (i) thread pool: a number of independent tasks are organized in a shared or distributed queues and a thread requests a new task from the task queue after it completes the previous task; (ii) pipeline parallel: groups of threads executing different stages in a software pipeline on an incoming stream of data, with task queues between pipeline stages and (iii) mapreduce: different threads independently executing ``map" functions on incoming data before synchronizing on the reduce thread.

We note that Bodytrack from PARSEC
and kmeans from Phoenix that are both
classified as barrier-based, actually use
mixed modes of parallelism: barriers 
across iterations, but thread pool and 
mapreduce parallelism within barriers, respectively. Other mapreduce benchmarks used in this paper have a single ``reduce" operation towards the end of execution.


Although TPEq is not designed keeping the characteristics of the 
benchmarks in Table~\ref{tab:workloads-other}
in mind, we nonetheless also compare CS and MaxBIPS with TPEq on these benchmark applications.
In fact, for the thread pool and mapreduce benchmarks, we expect MaxBIPS to perform the best. 
However, we
find that TPEq is, in fact, competitive with, and in some cases outperforms, MaxBIPS for these
benchmarks as well.

Sixteen parallel threads were used 
for each benchmark except Dedup and Ferret, which allow only 14 parallel threads (we note, however, Dedup and Ferret are not barrier synchronization based benchmarks, which are the main focus of this work). 
For 16 parallel threads, the PARSEC benchmarks also launch a $17^{th}$ ``initialization" 
thread that executes by itself on a core at the highest power and performance configuration. For a fair comparison, we report execution times starting from the time when parallel threads are first launched to the end of program execution.  

\section{Experimental Evaluation}\label{expresults}

We have compared TPEq with state-of-the-art techniques discussed in Section~\ref{comparitive}. 
We briefly describe our implementation of these techniques.


\noindent \textbf{Criticality Stacks (CS)~\cite{cs}:} Our CS implementation is faithful to the one reported in ~\cite{cs}. 
In every epoch, the thread with the highest criticality and above a threshold of $1.2$ is accelerated on the fastest core configuration, and all the other configurations are set to the highest homogeneous configuration that consumes the remaining power budget. This thread is accelerated until its criticality value becomes less than $0.8$ or another more critical thread above the threshold is found, in which case the new critical thread is accelerated. 
In \textit{addition}, for a fair comparison, we ensure that 
if there is any residual power at this point, 
the remaining threads are accelerated to highest possible configurations in the order of decreasing thread criticality.
This is the baseline approach over which we will compare TPEq. 
Although one can potentially devise more elaborate heuristics that look at the next most critical thread(s), we are not aware of any principled way to use CS for fine-grained optimization as enabled by TPEq.

\noindent \textbf{MaxBIPS~\cite{maxbips}:} MaxBIPS uses the same epoch length and predictors (power and performance) as TPEq. The sum-IPS optimization is performed using an off-the-shelf ILP solver in Matlab~\cite{MATLAB:2013}, and the solutions are fed back to Sniper. 
Since running an ILP solver would not be a practical solution in a real implementation, we also implemented \textbf{MaxBIPS-heuristic}, a polynomial time heuristic solver
for the MaxBIPS objective function.




\subsection{Power and Performance Prediction}
The TPEq predictors were trained offline on a small subset of five randomly 
chosen benchmarks (out of 24) with different input sets as used in the rest of the experiments.
In terms of CPI prediction, we observe a mean absolute error of $13.7\%$ over more
than 100,000 samples collected over all benchmarks. Of this, $5.7\%$ can be attributed directly to temporal errors
from last-value prediction. The rest of the error comes from predicting the CPI on one core configuration based 
on measurements on another core configuration. The PIE prediction mechanism~\cite{pie} reports similar errors of $9-13\%$ with only two (big and little) core configurations, while we have five. Further PIE only predicts CPI on a different core configuration for the current epoch, while we predict CPI for the \textit{next} epoch.   

The mean absolute error in power prediction is $4.43\%$, which is competitive with the errors reported in the state-of-the-art~\cite{bircher,contreras2005power,singh2009real}. It is important to note that although the CPI predictions are used for predicting power, the power prediction error is lower for two reasons: (i) some positive and negative error terms from CPI estimation of individual cores cancel out in total power and (ii) power also has a constant static component.
Because of the inaccuracy in power prediction, we observe that 
the power consumption occasionally exceeds the 80W budget, but the average 
overshoot is only $~3W$ for both TPEq and MaxBIPS, and slightly higher for CS.
Furthermore, power overshoots are short-lived and we observed only \textit{one} 
instance in all our experiments where the overshoot exceeds $3W$ for more than 
three successive epochs. In this (rare) event, a throttling mechanism kicks in 
to reduce power consumption. Prior work on proactive dynamic power management 
makes similar observations about overshoots~\cite{ma2011scalable,maxbips}.


\subsection{Results on Barrier Synchronization Based Benchmarks}

Figure~\ref{fig:tpeq_compare_hom_het}(a) compares the execution time of TPEq to the competing state-of-the-art techniques, MaxBIPS and CS for the benchmarks in Table~\ref{tab:workloads}. 
Also shown are the mean speed-ups (with CS as baseline) separately for the homogeneous (HO)
and heterogeneous (HT) benchmarks.

Several observations can be made: first, we observe that TPEq is the best performing 
technique for all but one benchmark (out of 15). TPEq is up to $23\%$ faster than CS and up to $15\%$ faster
than MaxBIPS. On average, TPEq is $5\%$ and $11\%$ faster than CS for homogeneous and heterogeneous
benchmarks, respectively. 
The speed-up of TPEq over CS is greater for heterogeneous 
benchmarks since these benchmarks feature \textit{both} IPC and instruction count heterogeneity, and provide greater 
opportunities for fine-grained optimization of core 
configurations.
It is instructive to note that the performance 
improvements of TPEq are over and above techniques 
that are already very competitive: CS has been shown
to improve over both AGETS and BIS (all coarse-grained 
optimization techniques since they only speed-up the 
most critical thread), while MaxBIPS is the only general, fine-grained technique that we are aware of.
In addition, our results are over a wide range of 
barrier synchronization benchmarks over three benchmark suites without any benchmark sub-setting. 

\noindent \textbf{Does Relative Instruction Count Prediction Help?}
Figure~\ref{fig:tpeq_compare_hom_het}(b) compares TPEq with a version, TPEq-W,
in which we do not perform relative instruction
count prediction and instead set all the weights, $w(i)$, to one.
Effectively, TPEq-W assumes that all threads execute the same number of instructions, 
and only account for IPC heterogeneity.

Note that although TPEq and TPEq-W are nearly identical for all homogeneous benchmarks (as expected, since 
all threads execute the same number of instructions), the speed-up of TPEq over TPEq-W is \textit{significant}
for heterogeneous benchmarks --- $15\%$ on average and up to $20\%$.

\vspace{-0.05in}
\begin{figure}[h]
\centering
\includegraphics[width=0.49\textwidth]{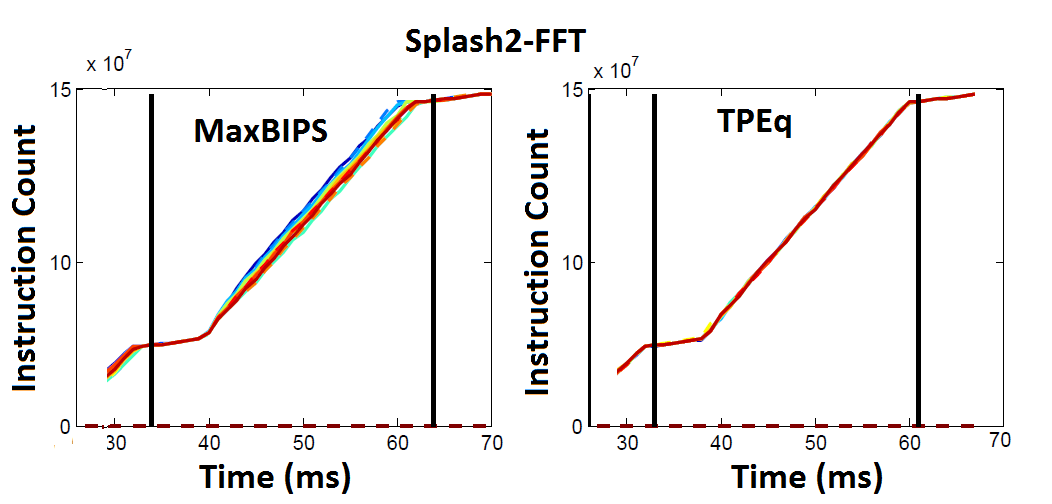}
\caption{Progress plots for FFT benchmark.}
\vspace{-0.1in}
\label{fig:pp_fft}
\end{figure}

\noindent \textbf{Why does TPEq outperform MaxBIPS and CS?}


To better understand why TPEq outperforms MaxBIPS, Figure~\ref{fig:pp_fft} shows
the MaxBIPS and TPEq progress plots for the FFT benchmark, focusing on
the second barrier phase. 
Note that, compared to the baseline
FFT progress plot in Figure~\ref{fi:ipc}, both the
MaxBIPS and TPEq progress plots have much less 
heterogeneity in thread progress. 
In fact, although MaxBIPS is not explicitly meant to equalize IPCs, we observe that
it many case, such as this, it speeds up low IPC threads and slows down high IPC
threads. 
Nonetheless, it is not able to equalize as IPCs as effectively as TPEq, 
as is clear from Figure~\ref{fig:pp_fft} 
--- the progress plots for all threads are almost perfectly aligned with TPEq, but more
spread out for MaxBIPS. Compared to the progress plot in Figure~\ref{fi:ipc}, TPEq
achieves an almost $60\%$ \textit{reduction in stall time}.

\begin{figure}
        \centering
                \includegraphics[width=0.5\textwidth]{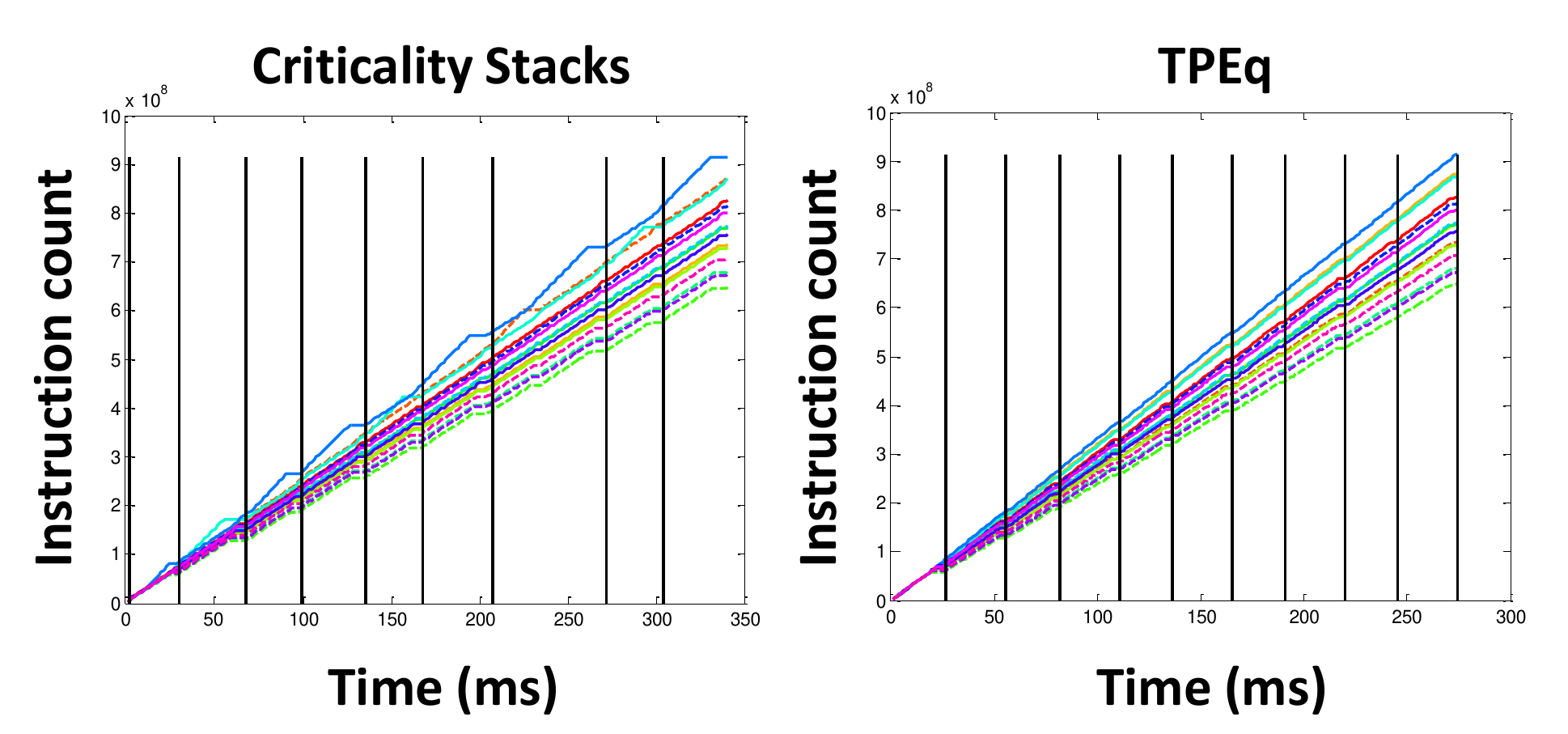}
        \caption{Progress plots for Fluidanimate using CS and TPEq.}\label{fig:pp_fa}
\end{figure}

\begin{figure}
\centering
\includegraphics[width=0.49\textwidth]{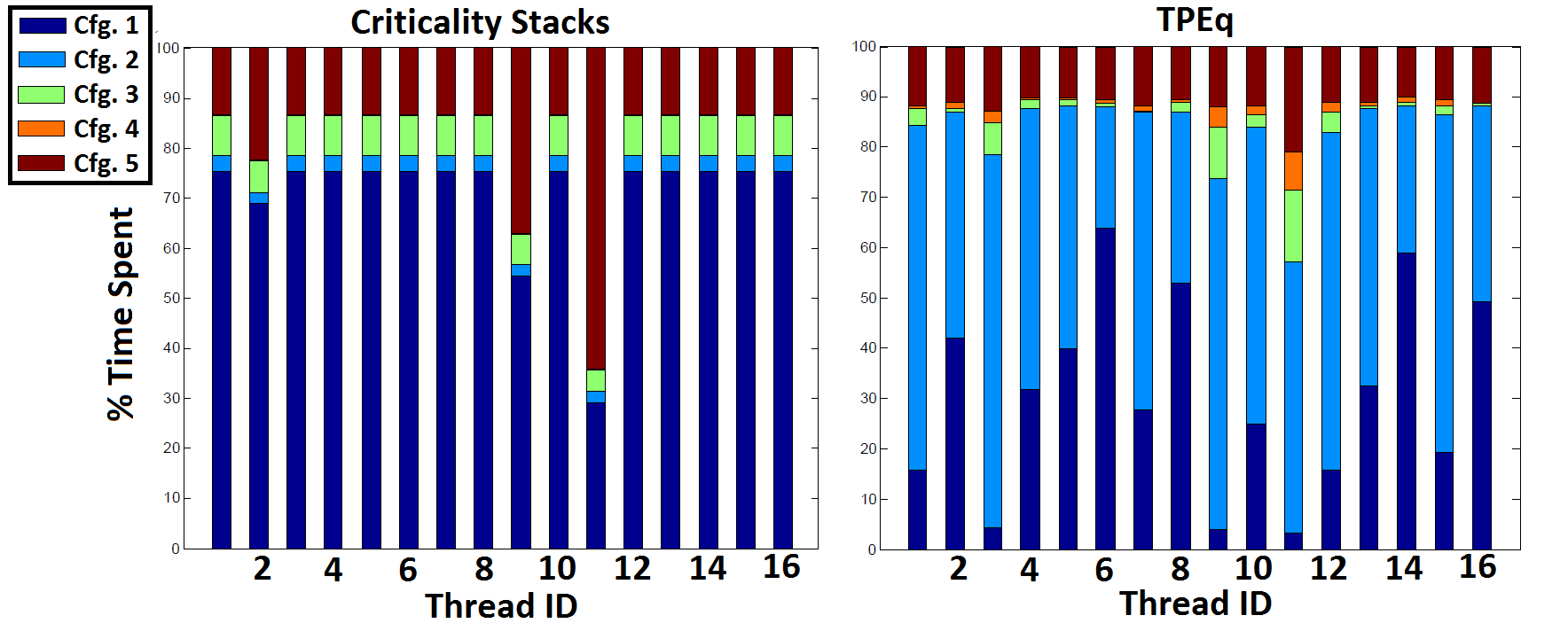}
\caption{Time spent by each thread in different configurations for CS and TPEq.}
\label{fig:timespent}
\end{figure}

Next we compare CS with TPEq, this time using Fluidanimate, a heterogeneous benchmark using progress plots shown in Figure~\ref{fig:pp_fa}. 
Again, observe that
TPEq is more successful in reducing thread stalls (regions where a thread's progress plot is flat) than CS, 
primarily because TPEq speeds-up or
slows-down each thread optimally so they reach barriers at (about) the same time, while
CS only speeds-up the most critical thread. 
In fact, the most critical thread identified by CS is sped-up more than necessary, and
end up stalling on the next barrier. 
Compared to the baseline, in which all threads are executed on identical cores within same power budget, 
TPEq reduces total stall time by as much as $50\%$, while CS only results in less than $20\%$
reduction in stall time.

Further insight can be obtained from Figure~\ref{fig:timespent}, 
which shows the time spent by each thread in each configuration 
for CS and TPEq. Although it is clear that both
both CS and TPEq identify Thread 11 as most critical (assigning it to higher power/performance configurations), TPEq assigns each thread (including Thread 11) to a great range of configurations since it 
is able to perform fine-grained optimization. In fact, configuration 4 is not utilized by CS at all, while this is not the case for TPEq.
%

\subsection{Results on Remaining Benchmarks}

\begin{figure}
  \centering
  \includegraphics[width=0.49\textwidth]{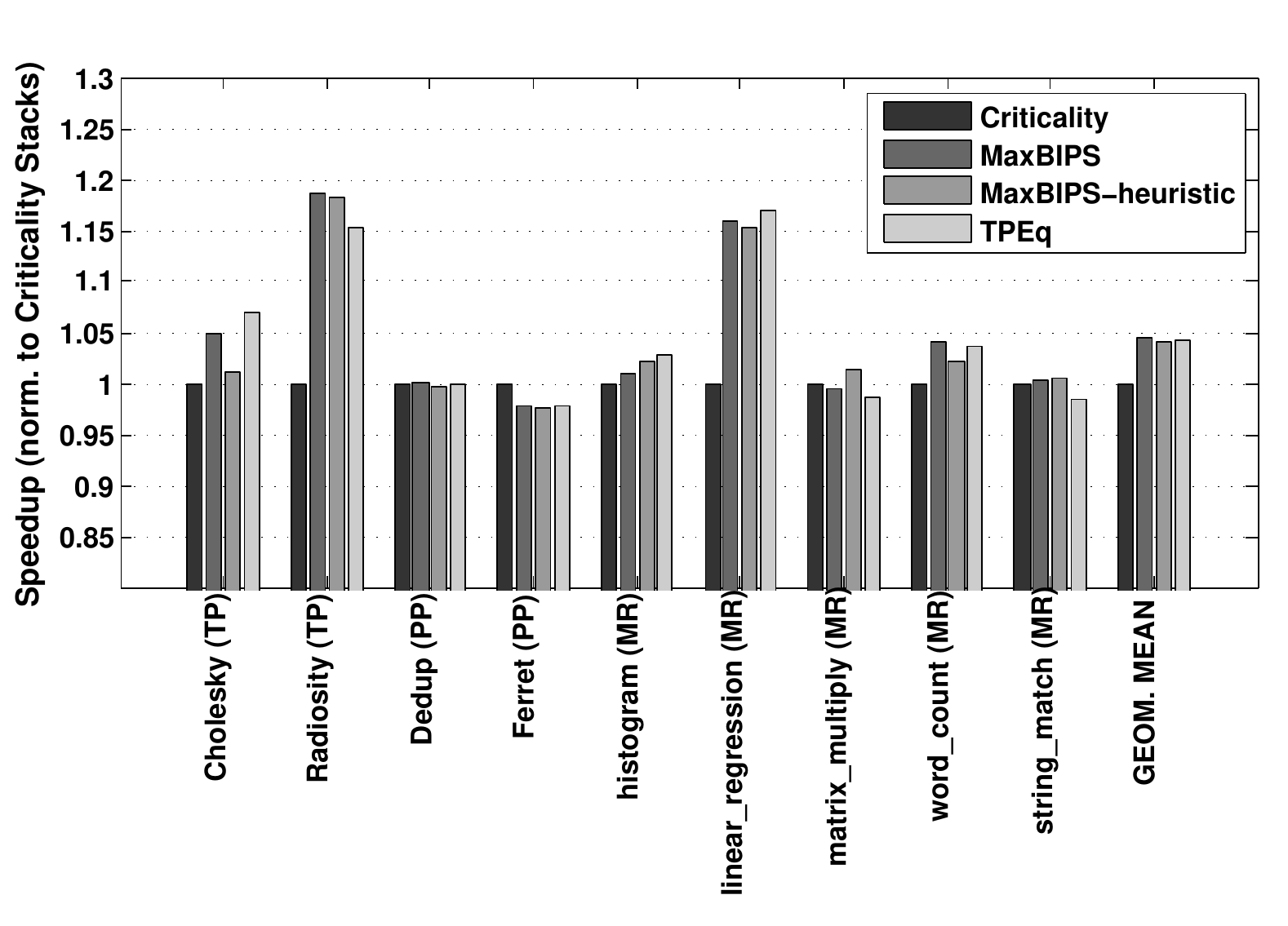}
  \vspace{-0.2in}
  \caption{Execution time on benchmarks in Table~\ref{tab:workloads-other}.}\label{fig:tpeq_compare_rem}
\end{figure}

Figure~\ref{fig:tpeq_compare_rem} shows the speed-up of TPEq and MaxBIPS
normalized to CS for the benchmarks in Table~\ref{tab:workloads-other} that
do not use barrier synchronization. We reiterate that TPEq is best suited
for barrier synchronization based parallel programs. However, since TPEq does
not explicitly look for barriers --- adaptation happens at regularly sized epochs
and threads asynchronously update their weights --- it can be used 
with any parallel program. 

We observe in Figure~\ref{fig:tpeq_compare_rem} that even for these benchmarks TPEq
outperforms CS. In addition, it is competitive with MaxBIPS on average and on a
per-benchmark basis. 
The improvement with respect to CS can be explained, in part, because TPEq (and MaxBIPS)
both perform fine-grained optimization while CS is coarse-grained. On the other
hand, the competitiveness of TPEq with MaxBIPS is more surprising since
MaxBIPS should be the ideal objective at least for the thread pool and mapreduce
benchmarks. We make the following observations to 
help explain the results: (a) for thread-pool and mapreduce benchmarks, 
we observed TPEq-W performs as well as TPEq and therefore TPEq is
primarily equalizing instruction counts in these settings, in other 
words, acting as a load balancer; and (b) for pipeline benchmarks we note
that the TPEq weights track thread criticality (to some extent), since 
the least (most) critical threads frequently (rarely) stall on full/empty queues. 
Nonetheless, we note that more work needs to be done on 
generalizing 
TPEq for other modes of parallelism beyond barrier synchronization.

\subsection{DVFS Results}

\begin{figure}
\centering
\includegraphics[width=0.4\textwidth]{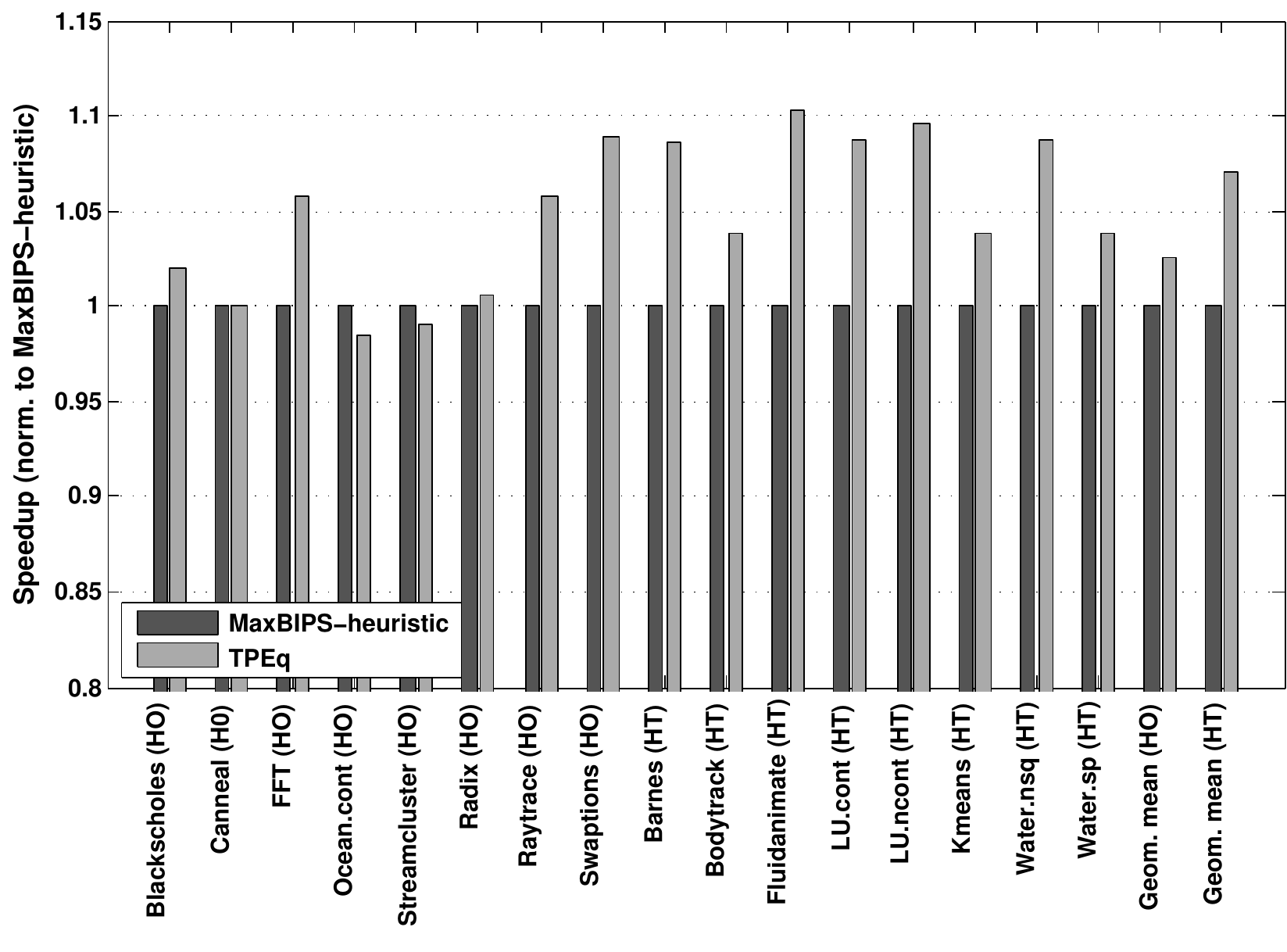}
\caption{Performance results for DVFS based dynamic adaptation comparing MaxBIPS with TPEq.}
\vspace{-0.1in}
\label{fi:dvfs_compare}
\end{figure}

TPEq can be easily modified for DVFS based dynamic adaptation. 
For DVFS, the 
power and CPI predictors are trained for every voltage-frequency configuration (as opposed to every 
micro-architectural configuration) using the 
model described in section~\ref{sec:prediction}. In addition, the second term of the 
progress metric
in Equation~\ref{eq:progress} is modified to $\frac{\mathcal{E} freq(i)}{w(i) CPI(i)}$, 
where $\mathcal{E}$ is now measured in seconds (as opposed to clock cycles) and $freq(i)$ 
is the frequency of thread $i$.
We performed DVFS experiments with five voltage-frequency levels ranging between \{0.8V, 2.5 GHz\} and 
\{1V, 3.5 GHz\} and compare the performance of TPEq with MaxBIPS in Figure~\ref{fi:dvfs_compare}.
The average performance improvement over MaxBIPS is ($6.9\%$) 
is slightly better than the improvements
over MaxBIPS obtained for micro-architectural adaptation, in part because of more accurate CPI prediction.



\subsection{TPEq Algorithm Runtime Overhead}

The TPEq optimizer needs CPI stack information from all cores
which happens implicitly via reads and writes to/from a shared
address space, along with required synchronization between threads (as discussed in Section~\ref{imp}). 
\begin{figure}
\centering
\includegraphics[width=0.4\textwidth]{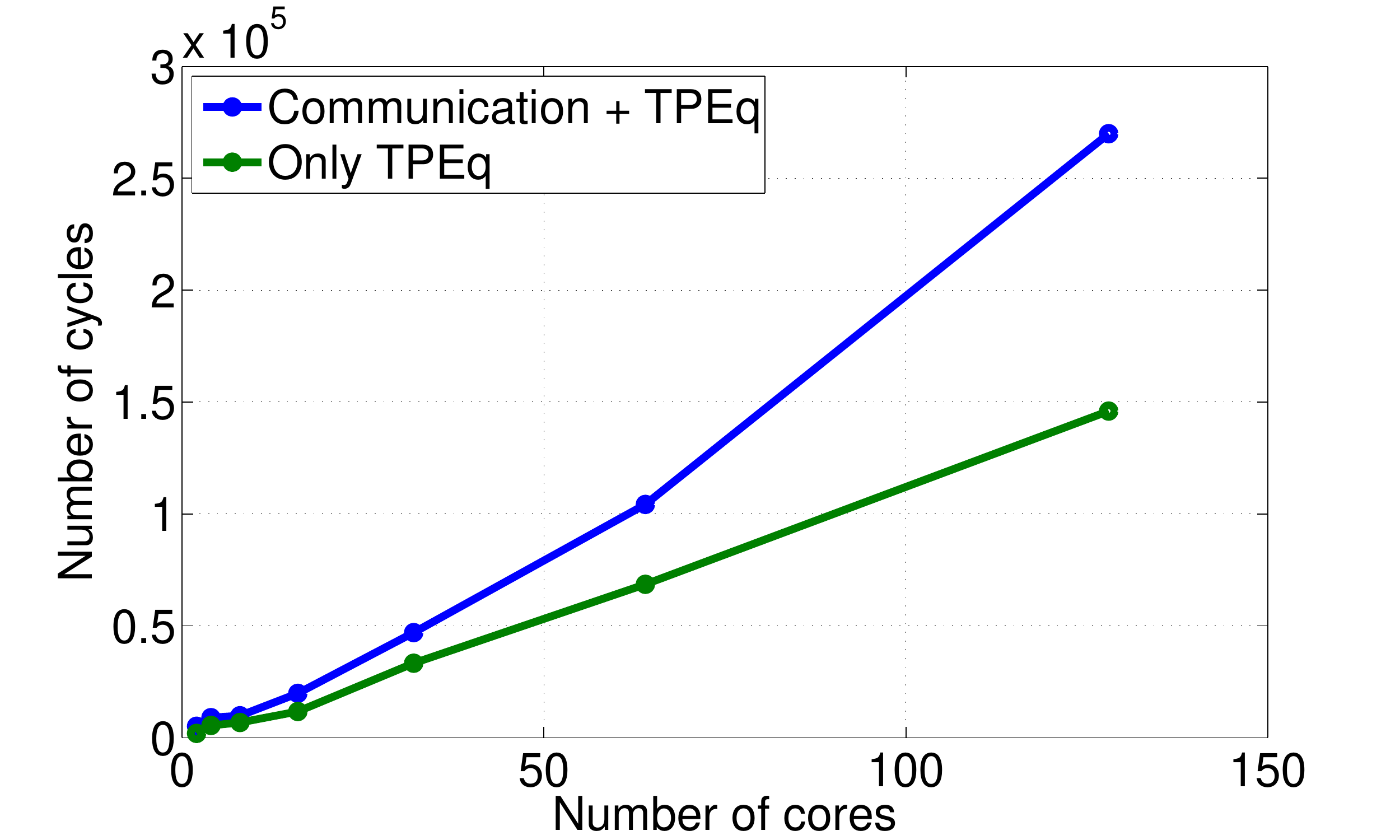}
\caption{TPEq run-time overhead}
\label{fi:time_overhead}
\end{figure}
For 16-cores, global communication takes roughly 10K cycles and the TPEq prediction and optimization procedures (including prediction overhead) 
take another 10K cycles. Together, the overhead amounts to $0.6\%$ for a 1 ms epoch length, assuming conservatively, 
that all other cores are stalled while the leader executes the TPEq routine. We also conduct a sensitivity analysis for TPEq overhead with increasing core/thread counts which is 
shown in Figure \ref{fi:time_overhead}. 
The close-to-linear scalability of TPEq optimization can be seen and is consistent with the complexity analysis of the algorithm which is $\mathcal{O}(MN \log N)$, with $N$ being the number of cores. 
We observe that the overhead of global communication grows faster than that of TPEq optimization. For many core systems with 100s of cores, hardware based communication and optimization support may be necessary. 

\section{Related Work}\vspace{-0.05in}\label{related}

Dynamic power and resource management of multi-core processors is an issue
of critical importance. 
Kumar et al.~\cite{kumar2003single} proposed the notion of single-ISA heterogeneous architectures to maximize 
power efficiency while addressing temporal and spatial application variations. Their focus was primarily on multiprogrammed workloads. A number of 
papers have proposed scalable thread scheduling and mapping techniques for such workloads~\cite{li2007efficient, teodorescu, shelepov2009hass, saez2010comprehensive, rangan2009thread}. 
Others have focused on leveraging asymmetry to 
increase the performance of multithreaded applications by identifying and accelerating critical sections~\cite{agets,cs,bis,uba, martonosi-tc}. 
A more recent work by Craeynest et. al~\cite{van2013fairness} proposes to use fairness-aware equal-progress scheduling on heterogeneous multi-cores, 
but it is unclear how this technique can be extended to optimal 
power-constrained performance maximization for adaptive multi-cores, 
which is the focus of this work.

The work on DVFS based dynamic adaptation of multi-core processors has made use of the sum-IPS/Watt~\cite{herbert2007} or MaxBIPS~\cite{maxbips} objectives, 
and different optimization algorithms including distributed optimization~\cite{ebi2009tape, sartori2009distributed} and 
control theory~\cite{ma2011scalable, hoffmann2011dynamic}. Cochran et al.~\cite{cochran2011pack} present a machine learning based approach based on offline workload characterization (and online prediction) but perform DVFS adaptation at a coarse time granularity of 100 billion uops.
Recently, Godycki~\cite{gody2014} et al. have proposed reconfigurable 
power distribution networks to enable fast, fine-grained, per-core voltage scaling and use this to \textit{reactively} (as opposed to TPEq's 
proactive approach) slow down
stalled threads and redistribute power to working threads. 
Also, unlike TPEq, this technique requires programmer inserted hints to 
determine the remaining work for each thread, and uses a heuristic 
approach to decide the voltage level of each core.

In the context of micro-architectural adaptation, ideas ranging from core-level to fine-grained 
power gating have been proposed~\cite{flicker, huang2003positional, ponomarev2001reducing, buyuktosunoglu2003energy,rcs}. 
Our work is most similar in spirit to \cite{liu2005exploiting}, which uses \emph{last-value predictor} at the barriers 
to set the frequency of cores so as to
save energy without compromising performance. 
However, TPEq is different from this technique on several counts. 
For one, \cite{liu2005exploiting} assumes that the slow-down
of each thread is directly proportional to frequency, while 
the TPEq optimizer is more 
general and works for complex power-performance relationships
that arise from micro-architectural adaptation and not just a simplified linear slow-down model. 
Second, TPEq does not require any explicit knowledge of a barrier event and is transparent to the programmer, while \cite{liu2005exploiting} requires programmer annotations.
Thus TPEq generalizes easily to a broader set of barrier synchronization based benchmarks, and
is not restricted to applications where barriers follow an easily
discernible template (i.e., outermost \emph{for} loops in OpenMP as studied by \cite{liu2005exploiting}). Finally, TPEq performs fine-grained adaptation (in time) at the granularity
of an epoch, while \cite{liu2005exploiting} only changes frequency once every
barrier phase.  

\section{Conclusion}
We proposed \textit{Thread Progress Equalization} (TPEq), a run-time mechanism to maximize performance under a power constraint for multithreaded applications running on multicores with support for fine grained dynamic adaption of core configurations. Compared with existing approaches, TPEq addresses all sources of inter-thread heterogeneity and determines in polynomial time the optimal configuration for each core so as to minimize execution time within a power budget.
Experimental results show that TPEq outperforms state-of-the-art techniques in the context of both micro-architecturally adaptive multicores, and while incurring modest execution time and hardware overheads. 





{\small                                                                                                                                                       
\bibliographystyle{ieee}
\bibliography{bibliography}
}






\end{document}